\documentclass{osa-article}
\usepackage{url}

\journal{osac}
\articletype{Research Article}
\begin{document}

\title{Spectrally Selective Solar Absorbers with High-temperature Insensitivity}

\author{Yanpei Tian,\authormark{1} Xiaojie Liu,\authormark{1} Alok Ghanekar,\authormark{2} and Yi Zheng\authormark{1,*}}

\address{\authormark{1}Department of Mechanical and Industrial Engineering, Northeastern University, Boston, MA 02115, USA\\
\authormark{2}Artech LLC, Morristown, NJ 07960, USA\\
}

\email{\authormark{*}y.zheng@northeastern.edu} %% email address is required



\begin{abstract}
It is of significance to incorporate spectral selectivity technology into solar thermal engineering, especially at high operational temperatures. This work demonstrates spectrally selective solar absorbers made of multilayer tungsten, silica, and alumina thin films that are angular insensitive and polarization-independent. An overall absorptance of 88.1\% at solar irradiance wavelength, a low emittance of 7.0\% at infrared thermal wavelength, and a high solar to heat efficiency of 79.9\% are identified. Additionally, it shows the annealed samples maintain an extremely high absorption in solar radiation regime over at least 600 $^{\circ}C$ and the solar absorbers after thermal annealing at 800 $^{\circ}C$ still work well in a CSP system with an even high concentration factor of over 100.
\end{abstract}

%%%%%%%%%%%%%%%%%%%%%%%%%%  body  %%%%%%%%%%%%%%%%%%%%%%%%%%
\section{Introduction}
The exploration of alternatives to traditional fossil fuels is invigorated over the past decades because of the energy crisis and increasing global warming, among which solar thermal energy is of intense interest due to its affluence and environmental sustainability. It gains much more attention recently in industrial heating, air condition, water desalination, and electricity generation \cite{kaempener2015solar,lim2017heat,kraemer2011high,kraemer2016concentrating,qiblawey2008solar}. However, it is strongly impeded to larger-scale engineering applications due to its low solar to heat conversion efficiency. Selective solar absorber, a key component that harvests solar radiations and converts it into thermal energy, greatly affects the thermal performance and efficiency of concentrated solar power (CSP), solar thermoelectric generator (STEG), and solar thermophotovoltaic (STPV) systems \cite{tian2018tunable,kraemer2011high,wang2012high,tian2019near}. Ideally, solar absorber should have highly sharp spectral selectivity, which means it possesses a unity absorptance and omnidirectional, polarization-independent nature \cite{yan2013metal,kajtar2016theoretical} in the solar irradiance range (visible and near-infrared regions) and shows no spontaneous thermal emittance losses in the mid-infrared regime due to blackbody radiation. The cut-off wavelength of a solar absorber, at which its absorptance spectrum changes sharply, is temperature-dependent owing to the wavelength shifting of blackbody radiation according to Wien's displacement law, so it is meaningful to investigate how the cut-off wavelength, the operational temperature, and solar concentration factors affect the energy conversion of CSP systems. Furthermore, an excellent high-temperature thermal stability is also highly desired for assuring solar absorbers to operate with high solar to heat conversion efficiency at fluctuated high temperatures. 

Numerous approaches have been discovered to obtain selective broadband absorbers, including both natural existing materials and micro/nanoscale patterned metamaterials. Natural materials based solar absorber, such as black carbon paint, black chrome \cite{pettit1982black,mcdonald1975spectral,sweet1984optical}, and Pyromark \cite{ho2014characterization} intrinsically exhibit high absorptance in the visible and near-infrared regions, as well as cermet \cite{ding2013self,gaouyat2014revealing,wackelgaard2015development,cheng2013improvement,zhang1998stainless,tang2016high}. However, they are not ideal for high-temperature applications, because they show high thermal leakage due to high emittance in mid-infrared wavelength regime. Moreover, the tunability of their spectral selectivity is relatively low, limiting their feasibility for diverse applications at different operational temperature or solar concentration factors. Besides natural materials, metamaterial with micro/nanoscale artificial structures displays spectral selectivity that cannot be achieved in naturally occurring materials \cite{liu2011metamaterials,ghanekar2015role}. The cut-off wavelength of high visible/near-infrared absorptance and low mid-infrared emittance can also be tuned through adjusting the geometric parameters of metamaterials. Plentiful selective metamaterials absorber, such as 1-D or 2-D surface gratings \cite{ghanekar2018optimal,wang2013perfect,lee2014wavelength,han2016broadband}, nanoparticles embedded dielectrics \cite{tian2019near,tian2018tunable,ghanekar2017mie,ghanekar2016novel}, cross-bar or nano-disk arrays \cite{chen2012dual,Nielsen2012Efficient}, and photonic crystals \cite{stelmakh2013high,celanovic2008two,rinnerbauer2013large,li2015large,jiang2016refractory} have been developed. However, these selective absorbers rely on complicated nanofabrication process, such as electron beam lithography (EBL) or focused ion beam milling (FIB), which is high-cost and time-consuming, which make it hard for large-scale industrial fabrication. The EBL and FIB techniques both rely on expensive equipment and materials supplies which make them stay on lab-scale fabrication and difficult to meet the industrial manufacturing requirements. Furthermore, the samples rely on the EBL and FIB techniques that are still not on the wafer-scale, which makes it time-consuming to expand to the industrial level. The fabrication process of photolithography is complicated and needs to be conducted in a cleanroom atmosphere. Compared with the abovementioned technique, vacuum deposition methods, either physical vapor deposition (PVD) or chemical vapor deposition (CVD), can fabricate large-scale samples and the cleanroom atmosphere is not mandatory in the industrial fabrication. The thermal stress of these structured based absorbers at elevated high temperatures will cause the unrecoverable damage of the surface topography and yields the loss of spectral selectivity. Hence, selective solar absorber that is lithography-free and holds high tunability of spectral selectivity and high-temperature stability are highly required in solar thermal energy engineering. 

Multilayer selective absorbers based on metal-insulator-metal (MIM) resonance have been demonstrated as an alternative approach to achieve wavelength selectivity \cite{chirumamilla2016multilayer,langlais2014high,nuru2014heavy} and can be fabricated by simple, cost/time effective, and large-scale vacuum deposition methods, such as sputtering, evaporation or chemical vapor deposition \cite{gordon2001solar}. However, the high-temperature stability of MIM resonance absorber coated with an anti-reflection layer because of possible thermal stress and material oxidation requires further investigations as well as its thermal performance at different operational temperatures and solar concentration factors. Refractory materials, such as tungsten, alumina, and silica (with bulk materials melting points of 3,422 $^{\circ}$C, 2,072 $^{\circ}$C, and 1,710 $^{\circ}$C, respectively) \cite{chirumamilla2016multilayer}, based multilayer broadband absorber can be a promising candidate with high-temperature stability. In addition, tungsten (W), alumina (Al$_2$O$_3$), and silica (SiO$_2$) have low thermal expansion coefficients of 4.2 $\times$ 10 $^{-6}$ m/(m$\cdot$K), 5.4 $\times$ 10 $^{-6}$ m/(m$\cdot$K), and 0.55 $\times$ 10 $^{-6}$ m/(m$\cdot$K) \cite{chirumamilla2016multilayer} at room temperature, respectively and these values are at the same order, which helps maintain the surface topography after thermal annealing though the thermal expansion coefficient is temperature-dependent. Additionally, tungsten is a good solar radiation absorber in the visible region since the real part of the dielectric permittivity is positive below 900 nm, among which over 50\% of solar radiation power distributes, and results straightly to high visible light absorption. Simultaneously, W shares the similar properties of common metals, such as silver, aluminum, and gold, that are highly reflective in the mid-infrared wavelength region. The Al$_2$O$_3$ and W thin layers arrange alternatively forming a MIM resonator and thereby exhibit enhanced absorptance of visible and near-infrared light. The thin SiO$_2$ layer on top of the MIM resonator serves as an anti-reflection layer to enhance the visible light absorption. Both Al$_2$O$_3$ and SiO$_2$ layers sandwich the easy-oxidized W layer and serve as a protective layer to ensure the high-temperature stability. 

In this work, we theoretically design and experimentally fabricated an ultrathin selective solar absorber based MIM resonance and SiO$_2$ anti-reflection effects. The angular and polarization-independent spectral reflectivity is identified. The reflection of both transverse electric (TE) and transverse magnetic (TM) polarizations remains low at incident angles of up to 85$^\circ$ within the visible and near-infrared region. High-temperature thermal treatment is further investigated to prove its thermal insensitivity. Scanning electron microscope (SEM) is employed to explore the topography variations after thermal annealing at different temperatures to shed light on how the topography of samples affects their spectral selectivity. The simulations of thermal performance for the fabricated absorber under a one-day sunlight cycle elucidates its potentiality in mid to high-temperature solar thermal energy systems.

\section{Fundamental theory}
\subsection{Energy conversion efficiency analysis of ideal solar absorber}

\begin{figure}[ht]
\centering
\includegraphics[width=0.45\textwidth]{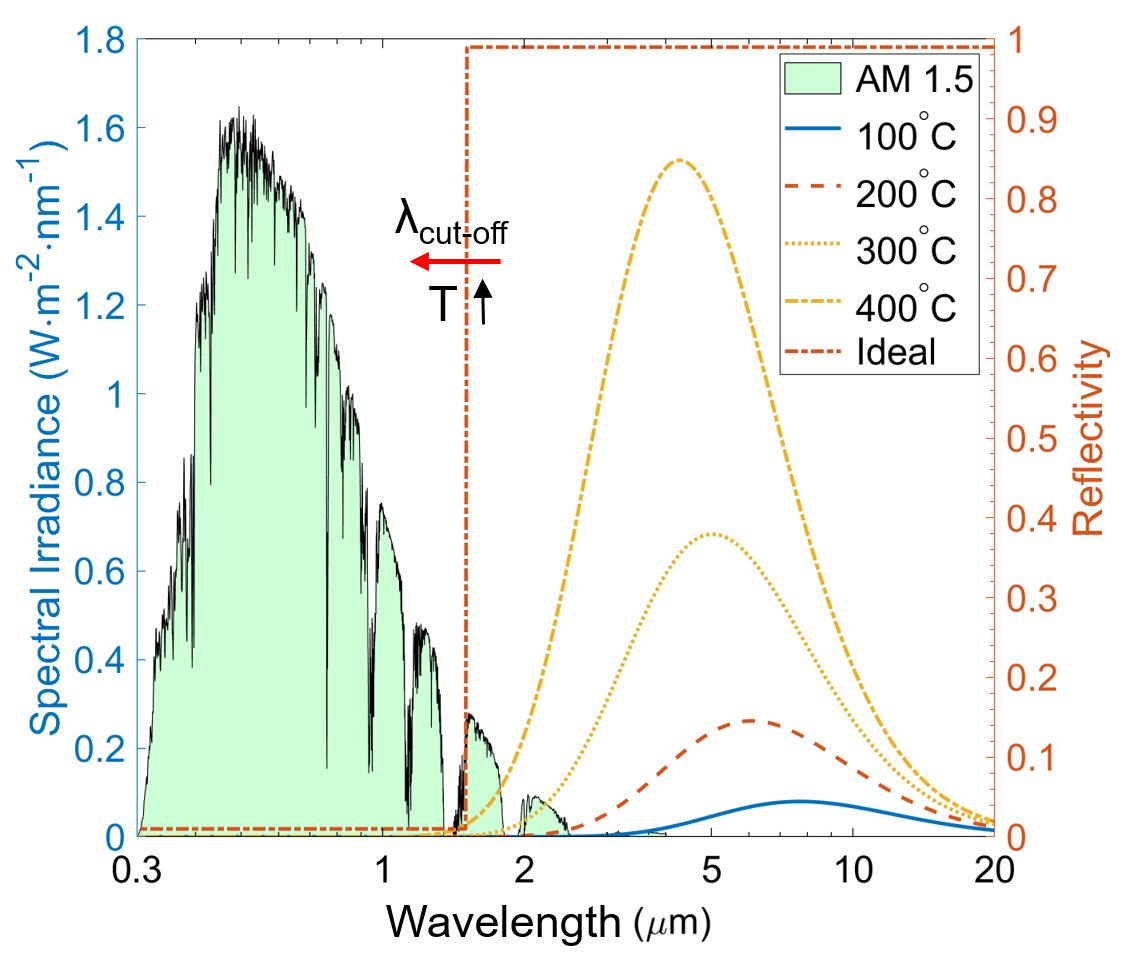}
\caption{ \label{fig:Fig_1} Solar spectral irradiance (AM 1.5, global tilt), radiative heat flux of blackbody thermal radiation, and reflectivity spectrum of ideal selective solar absorber.}
\end{figure}

To approach the perfect design of selective solar absorber, the affecting parameters of solar to heat conversion efficiency for the ideal selective solar absorber are theoretically investigated. Figure \ref{fig:Fig_1} shows the spectral irradiance distributions of solar radiative heat flux, as well as the thermal radiative heat flux of a blackbody at various temperatures. It can be found that most of the solar irradiance is distributed in the visible and near-infrared region (0.3 $\mu$m $\sim$ 2.5 $\mu$m), while most of the thermal radiation from a blackbody spreads within the mid-infrared regime according to Wien's displacement law. When a blackbody is at 400 $^{\circ}C$, the thermal emission is at the same order of solar heat flux. Therefore, it is crucial to maximizing the absorption of solar radiation and minimizing the energy loss from the thermal re-emission in the mid-infrared spectral regime. Additionally, the distribution wavelength regime of blackbody radiation moves to a shorter wavelength as its temperature increases, as illustrated by thermal radiation curves of the blackbody at 100 $^{\circ}C$, 200 $^{\circ}C$, 300 $^{\circ}C$, and 400 $^{\circ}C$ in Fig. \ref{fig:Fig_1}. The cut-off wavelength, $\lambda_\text{cut-off}$ shifts to a shorter wavelength so as to reduce the blackbody thermal re-emission to approach a maximum efficiency. Figure \ref{fig:Fig_1} also exhibits the reflectivity spectrum of an ideal solar absorber, which shows zero reflectivity below $\lambda_\text{cut-off}$ to increase the absorption of solar radiation and unity reflectivity beyond the $\lambda_\text{cut-off}$ to reduce the thermal emission. Moreover, the principle differs between unconcentrated and concentrated solar power applications, considering that the concentration factor of a focusing lens can be designed up to several thousands of times for now. Note that, the operational temperature of the solar absorber varies greatly when the circulation rate of working fluids or thermal loads differs. The energy conversion efficiency of solar absorber alters as a function of the cut-off wavelength, the concentration factor of solar light, and the operational temperature of the solar absorber. Hence, it is a trade-off to design a nearly perfect selective solar absorber for diverse engineering applications. 

The solar to heat conversion efficiency of solar absorber is defined by this following equation, assuming no heating conduction and convection losses: 
\begin{equation}
\label{eq:e1}
\eta=\frac{\alpha_{\text{abs}} CF \cdot Q-\epsilon_{\text {abs}}\left(\sigma T_\text{abs}^{4}-\sigma T_{\text {amb}}^{4}\right)}{CF \cdot Q}
\end{equation}
where $CF$ is the concentration factors, $Q$ is the solar radiative heat flux at AM 1.5 (global tilt) \cite{solaram1_5}. $\sigma$ is the Stefan-Boltzmann constant.  $T_\text{abs}$ is the operational temperature of absorber, and $T_\text{amb}$ is set to be 25 $^{\circ}C$ as the ambient temperature. $\alpha_\text{abs}$ and $\epsilon_\text{abs}$ is the total absorptance and emittance of the solar absorber, respectively, which are expressed as the following:

\begin{equation}
\label{eq:e2}
\alpha_{\text{abs}}=\int_{0}^{\infty} \alpha_{\lambda,\text{abs}}^{\prime} I_{\mathrm{AM} 1.5}(\lambda) d \lambda / \int_{0}^{\infty} I_{\mathrm{AM} 1.5}(\lambda) d \lambda
\end{equation}

\begin{equation}
\label{eq:e3}
\epsilon_{\text{abs}}=\int_{0}^{\infty} \epsilon_{\lambda,\text{abs}}^{\prime} I_{\mathrm{BB}}\left(\lambda,T_\text{abs}\right) d \lambda / \int_{0}^{\infty} I_{\mathrm{BB}}\left(\lambda,T_\text{abs}\right) d \lambda
\end{equation}
where $\alpha_{\lambda,\text{abs}}^{\prime}$ and $\epsilon_{\lambda,\text{abs}}^{\prime}$ are the wavelength-dependent absorptance and emittance for the solar absorber at the room temperature, respectively. $I_{\mathrm{AM} 1.5}(\lambda)$ is the spectral irradiance intensity of solar radiation at AM 1.5 and $I_{\mathrm{BB}}\left(\lambda,T_\text{abs}\right)$ is the spectral blackbody radiative intensity at a certain operational temperature, $T_\text{abs}$. For the theoretical calculation of $\alpha_{\lambda,\text{abs}}^{\prime}$ and $\epsilon_{\lambda,\text{abs}}^{\prime}$, the spectral integration range is limited within 0.25 $\mu$m $\sim$ 4.0 $\mu$m, where the available AM 1.5 data covers.  

\begin{figure}[ht]
\centering
\includegraphics[width=1\textwidth]{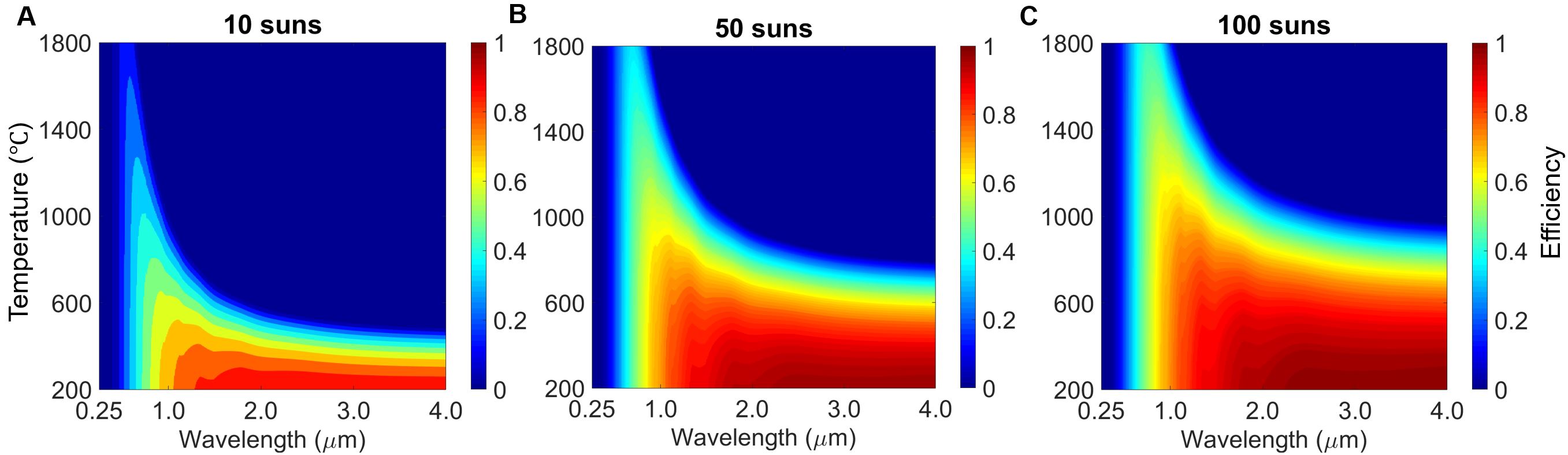}
\caption{\label{fig:Fig_2} (\textbf{A}), (\textbf{B}), and (\textbf{C}) Solar to heat energy conversion efficiency of solar absorber contour plotted against the absorber operational temperature and cut-off wavelength at different solar concentration factors of 10, 50, and 100, respectively.
}
\end{figure}

Figure \ref{fig:Fig_2} illustrates the energy conversion efficiency, $\eta_\text{abs}$, as a function of the absorber operational temperature, $T_\text{abs}$, and the cut-off wavelength, $\lambda_\text{abs,cut-off}$ at different concentration factors, $CF$, 10 suns, 50 suns, and 100 suns, in which the red color represents higher efficiency, and the blue means lower efficiency. Note that, these energy efficiency contour plots are obtained by analyzing Eqs. (\ref{eq:e1}), (\ref{eq:e2}), and (\ref{eq:e3}). It can be observed in Figs. \ref{fig:Fig_2}\textbf{A}, \ref{fig:Fig_2}\textbf{B}, and \ref{fig:Fig_2}\textbf{C} that the efficiency curve shares the similar potential of increasing to maximum from zero and then decreasing as $\lambda_\text{cut-off}$ sweeps from 0.25 $\mu$m to 4.0 $\mu$m. For the absorber with a temperature of 600 $^{\circ}C$ under 50 suns, the efficiency is 18.9\% at $\lambda_\text{cut-off}$ of 0.5 $\mu$m, and increases to a maximum of 94.3\% at $\lambda_\text{cut-off}$ of 1.8 $\mu$m, then decreases to 72.8\% at $\lambda_\text{cut-off}$ of 4 $\mu$m. According to Wien's displacement law, the thermal radiative heat flux of 600 $^{\circ}C$ blackbody is distributed with 1 $\mu$m to 16 $\mu$m and reaches its maximum at 3.2 $\mu$m. Therefore, the solar radiation matters at a shorter wavelength (0.25 $\mu$m to 1.8 $\mu$m), while the thermal losses due to blackbody re-emission become prominent at a longer wavelength (1.8 $\mu$m to 4.0 $\mu$m). The maximum efficiency decreases and the corresponding $\lambda_\text{cut-off}$ shifts to a shorter wavelength at 50 suns as the temperature increases. The energy conversion efficiency at other concentration factors follows the same rules, as can be seen in Fig. \ref{fig:Fig_2}. The maximum efficiency under 50 suns happens at 1.8 $\mu$m with 94.3\% at 600 $^{\circ}C$, at 1.3 $\mu$m with 79.6\% at 1,000 $^{\circ}C$, and at 0.8 $\mu$m with 47.9\% at 1500 $^{\circ}C$. It is reasonable that $\lambda_\text{cut-off}$ shifts to left, since the main distribution wavelength of blackbody radiation moves to a shorter wavelength as the temperature increases. The corresponding maximum efficiency decrease from 94.3\% to 47.9\% as the temperature increases from 600 $^{\circ}C$ to 1500 $^{\circ}C$, since the overlap between the distribution area of the solar irradiance and blackbody radiation becomes larger, considering that the wavelength region of solar radiation keeps fixed, while the thermal radiation regime of the blackbody shifts to the left and the magnitude of thermal radiative heat flux increases along with the increasing of absorber temperature. The maximum efficiency of the absorber at 800 $^{\circ}C$ happens at 1.3 $\mu$m with 81.3\% under 10 suns, at 1.4 $\mu$m with 86.9\% at 50 suns, and at 1.8 $\mu$m with 91.1\% at 100 suns. It can be found that the $\lambda_\text{cut-off}$ moves to a longer wavelength and the maximum efficiency increases as well when the concentration factors increase from 10 suns to 100 suns. It is easy to understand that the absorbed heat flux of solar power increases while the distribution wavelength regime and the radiation rate of blackbody keep fixed at a certain temperature, and the solar radiation becomes dominant at a high $CF$.%, which can be seen in the Eq. \ref{eq:e2}.

\begin{figure}[!t]
\centering
\includegraphics[width=0.9\textwidth]{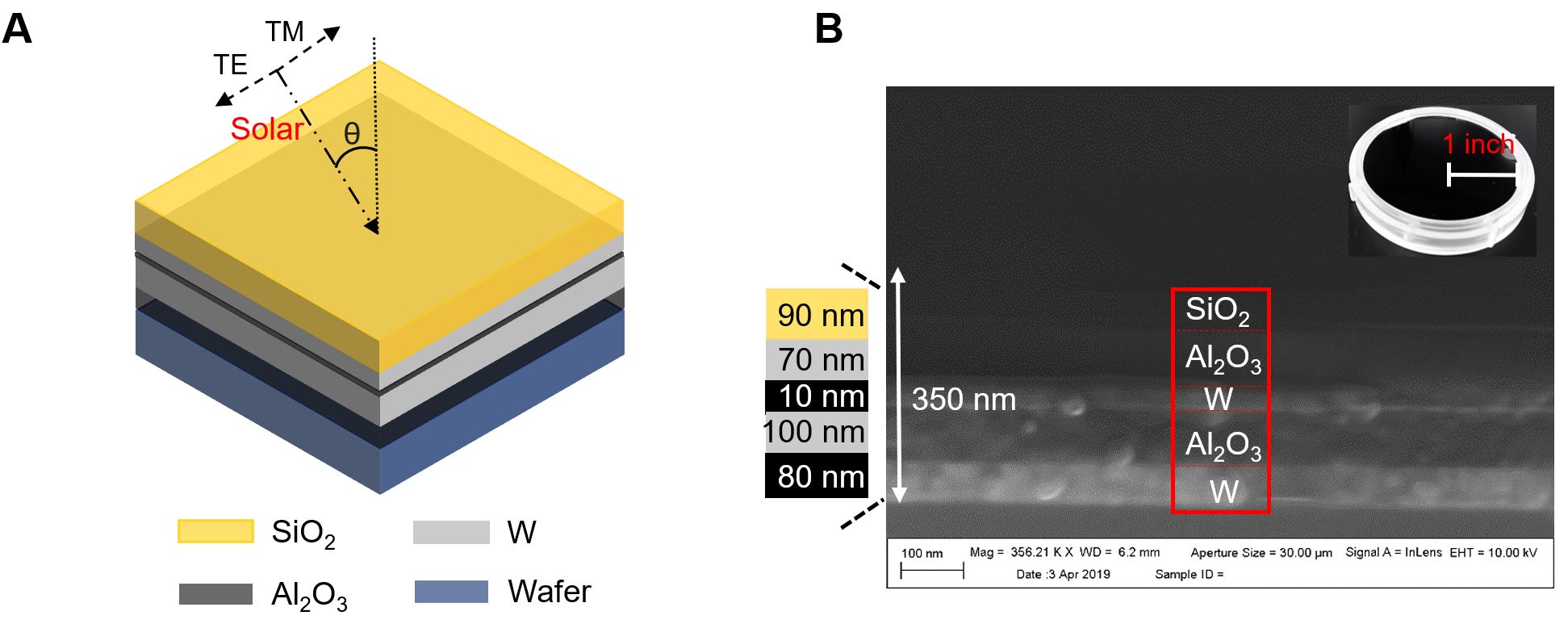}
\caption{\label{fig:Fig_3} (\textbf{A}) 3-D schematic of a multilayer stack consisting of W, Al$_2$O$_3$, and SiO$_2$. The incidence angle, $\theta$, is defined as the angle between solar incident radiation and the surface normal. (\textbf{B}) A cross-section SEM micrograph of the fabricated sample, the 2-D schematic shows the thickness of each layer for the multilayer stack, and the insect is an image of sample fabricated on a 2-inch silicon wafer staying in a wafer carrier case.       
}
\end{figure}

\begin{table}[ht]
\small
  \caption{\ Calculated energy conversion efficiency of the designed solar absorber at different operational temperature under 100 suns}
  \label{Tab1}
  \begin{tabular*}{1.0\textwidth}{@{\extracolsep{\fill}}lllll}
    \hline
    Temperature ($T_\text{abs}$, $^{\circ}C$ ) & 200 & 400 & 500 & 600 \\
    \hline
    Efficiency ($\eta$, \%) & 81.0 & 82.3 & 81.5 & 79.8 \\
    \hline
  \end{tabular*}
\end{table}

%\subsection{Sample Characterizations}
\section{Results}
\subsection{Design and spectrometric characterizations of absorber samples}

The thickness for different layer stacks of the solar absorber is optimized using MATLAB optimizations toolbox. The proposed planer structure is fabricated by employing the magnetic sputtering technique. Figure \ref{fig:Fig_3}\textbf{A} illustrates the 3-D schematic of proposed multilayer stack consisting of five layers (i.e., SiO$_2$, Al$_2$O$_3$, W, Al$_2$O$_3$ and W from top to bottom, successively). The thickness of each layer for the multilayer stack is shown in the 2-D schematic of Fig. \ref{fig:Fig_3}\textbf{B}. The thickness of different layers for the fabricated sample is confirmed by the cross-section view of FE-SEM shown in Fig. \ref{fig:Fig_3}\textbf{B}. Considering the thickness of the bottom W layer, it is reasonable to consider the sample is opaque at the wavelength of interest and the solar absorptance of the absorber, $\alpha_\text{abs}$ = 1 $-$ $\rho_\text{abs}$. 
% Figure \ref{fig:Fig_5}\textbf{A} shows the simulated spectral reflectivity for designed selective solar absorber with the thickness of each layer optimized at 100 suns. 
% It is observed that the designed solar absorber exhibits a spectral selectivity with solar reflectivity $\rho_\text{abs}$ $<$ 0.05 in the solar spectrum and thermal reflectively $\rho_\text{abs}$ $>$ 0.95 in the infrared region.
Table \ref{Tab1} shows the calculated energy conversion efficiency of the designed solar absorber under 100 suns calculated using the simulated reflectivity spectrum at different temperatures. The solar to heat efficiency of absorber reaches a maximum at the temperature of 400 $^{\circ}C$.

\subsection{Diffuse and polarization-independent behaviors of designed solar absorbers}
Generally, it is significant to keep the solar absorber face to the sun all the time to increase the solar conversion efficiency, therefore, a solar tracker system becomes a key component in the CSP engineering application though it consumes the self-generated electricity. It is urgent to avoid the additional energy consumption from the solar tracker system and develop a diffuse-like solar absorber that has an angular-independent spectral absorptance. 
Besides, polarization independence is also crucial for a perfect absorber to maximize solar energy absorption since solar radiation is randomly polarized. 

\begin{figure}[!t]
\centering
\includegraphics[width=1\textwidth]{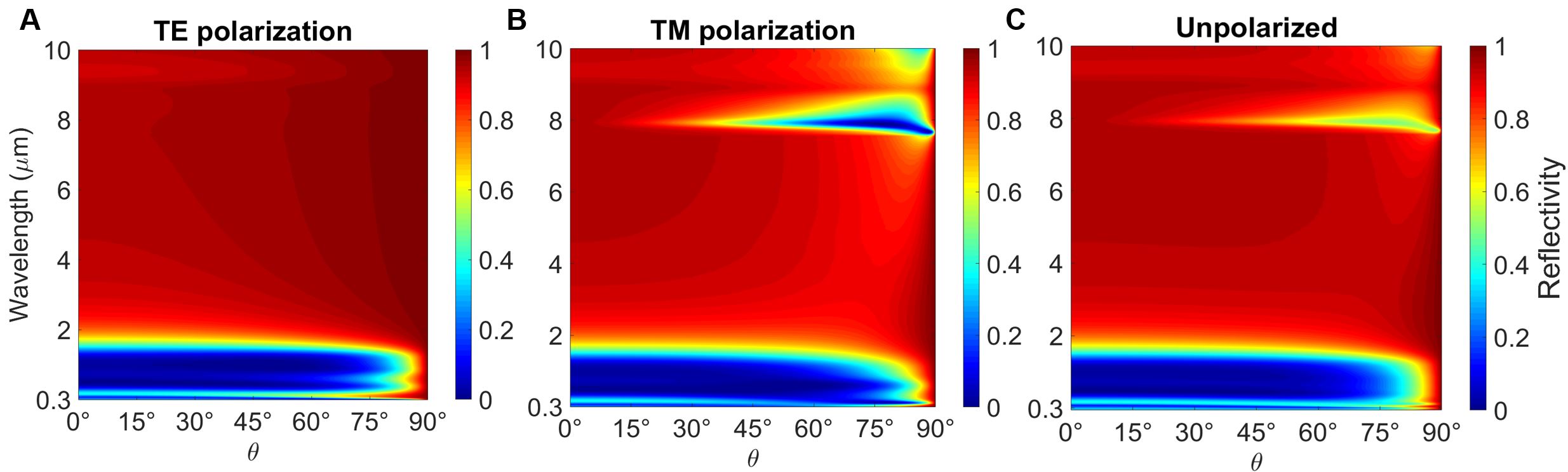}
\caption{ \label{fig:Fig_4} (\textbf{A}), (\textbf{B}), and (\textbf{C}) Angle dependent TE polarization, TM polarization, and unpolarized reflectivity of selective solar absorber contour plotted against wavelength and angle of incidence, $\theta$.  
}
\end{figure}

Here, Figure \ref{fig:Fig_4} manifests the simulated contour plot of the spectral reflectivity for the multilayer solar absorber as a function of incident angle, $\theta$, and wavelength at TE, TM polarized waves and unpolarized waves, in which the red color means higher reflectivity, while the blue color represents lower reflectivity. It is exhibited that the reflectivity remains lower than 0.1 with visible and near-infrared (from 0.25 $\mu$m to 1.7 $\mu$m) in Figs. \ref{fig:Fig_4}\textbf{A} and \ref{fig:Fig_4}\textbf{B}. At a wavelength, $\lambda$ = 0.55 $\mu$m in which the irradiation peak of solar lies, the reflectivity is 0.129 at normal incidence, and it decreases slightly to 0.117 at $\theta$ = 30$^{\circ}$, then reduces to 0.109 at $\theta$ = 60$^{\circ}$ for TE waves. Even at $\theta$ = 75$^{\circ}$, the reflectivity is as low as 0.179 for TE waves. For TM waves, the reflectivity is 0.129 at normal incidence, and it decreases slightly to 0.093 at $\theta$ = 30$^{\circ}$, then reduces to 0.044 at $\theta$ = 60$^{\circ}$. The reflectivity maintains at 0.113 even at $\theta$ = 75$^{\circ}$ for TM waves at $\lambda$ = 0.55 $\mu$m. The reflectivity across the infrared region (from 0.8 $\mu$m to 1.7 $\mu$m) follows similar rules. It proves the multilayer solar absorber remains low reflectivity over a broad spectral range for both polarizations. The angular-insensitivity spectral reflectivity for unpolarized waves is shown in Fig. \ref{fig:Fig_4}\textbf{C}. It can be seen that the reflectivity is insensitive to the incident angle over a large range, and low reflectivity exists over the visible and near-infrared region. Note that, although the reflectivity shows a dip at $\lambda$ = 8 $\mu$m, the peak wavelength of 400$^{\circ}C$ blackbody thermal radiation lies at 3.2 $\mu$m. For higher temperature applications, the peak wavelength shows blue shifts to a shorter wavelength, so it is reasonable that this dip does not affect the thermal performance of the designed solar absorber. The overall absorptance of the fabricated samples at solar irradiance wavelength regime is 87.19\%, while it shows a low emittance of 7.13\% at a thermal wavelength with the measured spectral reflectivity, assuming that the operational temperature of absorbers is 100$^{\circ}C$.

\begin{figure}[!b]
\centering
\includegraphics[width=1\textwidth]{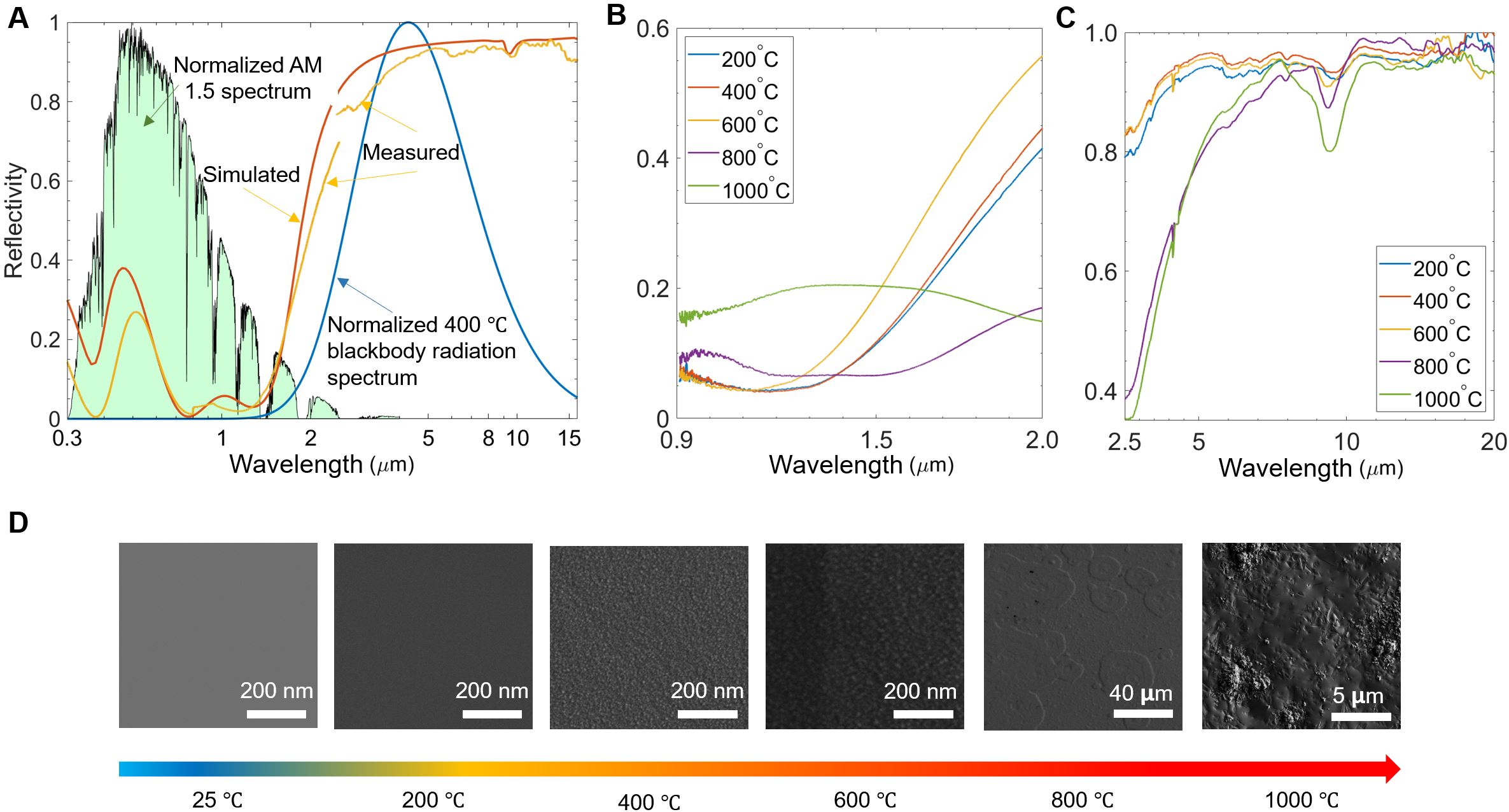}
\caption{\label{fig:Fig_5}(\textbf{A}) Normalized spectral distribution for radiative heat flux of solar (AM 1.5) and blackbody thermal radiation (400 $^{\circ}C$), as well as the simulated and measured reflectivity spectra of a multilayer solar absorber; (\textbf{B}) Near-infrared and (\textbf{C}) mid-infrared  spectral reflectivity of the multilayer absorber measured by FTIR spectrometer at room temperature (25 $^{\circ}C$) after thermal treatment at 200 $^{\circ}C$, 400 $^{\circ}C$, 600 $^{\circ}C$, 800 $^{\circ}C$, and 1,000 $^{\circ}C$ for 6 hours. (\textbf{D}) SEM topographic images of the fabricated multilayer solar absorber sample after thermal annealing at 200 $^{\circ}C$, 400 $^{\circ}C$, 600 $^{\circ}C$, 800 $^{\circ}C$, and 1,000 $^{\circ}C$ for 6 hours.
}
\end{figure}

\subsection{Spectral selectivity of the fabricated sample at room temperature}
The hemispherical reflectivity of the fabricated sample is characterized by UV/VIS/NIR spectrophotometer in the visible and near-infrared region (from 0.3 $\mu$m to 2.5 $\mu$m) and FTIR spectrometer in the mid-infrared regime (from 2.5 $\mu$m to 16 $\mu$m) at room temperature. The measured spectral reflectivity breaks at 2.5 $\mu$m since the incident angle of the light beam for these two spectrometers is different and the measurement mechanism of embedded detector differs, as discussed in the following method section. The simulated spectral reflectivity is also interrupted at 2.5 $\mu$m to match the measured spectrum. The simulation of reflectivity spectrum is executed at $\theta$ = 0$^{\circ}$ and 12$^{\circ}$ at visible/near-infrared and mid-infrared regime, respectively. A good match between the simulated and measured spectra is seen in the near-infrared and mid-infrared spectral region, while the measured reflectance is lower from 0.3 $\mu$m to 0.5 $\mu$m, which is highly desired for enhanced solar radiation absorption. 

\subsection{Thermal insensitivity test and thermal failure mechanism from SEM characterizations}

\begin{table}[b]
\small
  \caption{\ Overall thermal absorptance and thermal emittance as fabricated and after thermal annealing for 6 hours at different temperatures ($T_\text{abs}$= 100$^{\circ}C$)}
  \label{Tab2}
  \centering
  \begin{tabular*}{0.6\textwidth}{@{\extracolsep{\fill}}lll}
    \hline
    Temperature ($^{\circ}$C) & $\alpha_\text{abs}$ & $\epsilon_\text{abs}$\\
    \hline
     25 & 0.872 & 0.071 \\
     200 & 0.843 & 0.082 \\
     400 & 0.838 & 0.063 \\
     600 & 0.809 & 0.160 \\
     800 & 0.892 & 0.190 \\
     1,000 & 0.812 & 0.211 \\
    \hline
  \end{tabular*}
\end{table}

Consistent spectral performance of the selective solar absorbers at high temperatures is significant, especially for CSP systems to maintain a high conversion efficiency under concentrated solar radiation. To evaluate the radiative properties of the absorber at different temperatures, we use the FTIR spectrometer, which covers 0.9 $\mu$m to 15 $\mu$m together with a gold integrating sphere, to measure the hemispheric reflectivity after thermal annealing at various temperatures for 6 hours. Figure \ref{fig:Fig_5}\textbf{B} and \textbf{C} exhibits the spectral reflectivity of the solar absorber after 6 hours thermal annealing at various temperatures. It can be seen that the spectral reflectivity of the tested samples barely changes from 200 $^{\circ}C$ to 600 $^{\circ}C$, indicating its excellent high-temperature stability. The near-infrared reflectivity starts to increase and the mid-infrared reflectivity begins to decrease dramatically when the annealing temperature is increased to 800 $^{\circ}C$ and 1,000 $^{\circ}C$. It indicates that spectral selectivity starts to fail possibly due to physical or chemical damages at high temperatures. Table \ref{Tab2} lists the overall thermal absorptance and thermal emittance as fabricated and after thermal annealing for 6 hours at different temperatures by integrating the wavelength from 0.9 $\mu$m to 15 $\mu$m. It is shown that both thermal absorptance, $\alpha_\text{abs}$, and emittance, $\epsilon_\text{abs}$, changes greatly and $\alpha_\text{abs}$ shows a potential of increase after 800 $^{\circ}C$ and 1,000 $^{\circ}C$ annealing, which is also displayed in Fig. \ref{fig:Fig_5}\textbf{B} and \textbf{C} that the cut-off wavelength begins to red-shift to a longer wavelength. Simultaneously, $\epsilon_\text{abs}$ increases with increasing of thermal annealing temperature, which indicates that the selectivity of the fabricated absorber becomes weaker because of the damages from high-temperature annealing. However, the fabricated samples still keep relatively low reflectivity (around 0.1 after 800 $^{\circ}C$ annealing, and around 0.2 after 1,000 $^{\circ}C$ annealing) in solar radiation region, and the reflectivity between 2.5 $\mu$m and 7.0 $\mu$m maintains at a low value. It indicates that the spectral selectivity has failed partially, but the solar absorber still works in a CSP system with a high $CF$ (over 100) which dominates the energy conversion efficiency.

To illustrate the mechanism that causes the degradation at 800 $^{\circ}C$, the samples are characterized under FE-SEM (SIGMA VP) before and after thermal annealing. Figure \ref{fig:Fig_5}\textbf{D} shows the SEM topographic images of the sample surface before and after being heated from 200 $^{\circ}C$ to 1,000 $^{\circ}C$ for 6 hours. From 25 $^{\circ}C$ to 200 $^{\circ}C$, it shows no apparent changes. When the temperature keeps going up from 400 $^{\circ}C$ to 600 $^{\circ}C$, the granulated protrusions with a diameter of about 20 nm show up at the samples surface, which can also be demonstrated in the reflectivity spectrum as seen in Fig. \ref{fig:Fig_5}\textbf{B} and \textbf{C}. However, when the absorber is further heated to 800 $^{\circ}C$, blisters with diameters around 40 $\mu$m are formed and cracks appear at 1,000 $^{\circ}C$, possibly due to the thermal press arising from the difference of thermal expansion coefficients of SiO$_2$ and Al$_2$O$_3$. Here, it leaves room for an improvement to approach to a prefect solar absorber by employing other materials with more similar thermal expansion coefficients with Al$_2$O$_3$ as an anti-reflection coating to avoid the thermal stress.

\section{Discussion}
\subsection{Solar conversion efficiency calculation for fabricated multilayer solar absorber}

\begin{figure}[!t]
\centering
\includegraphics[width=0.9\textwidth]{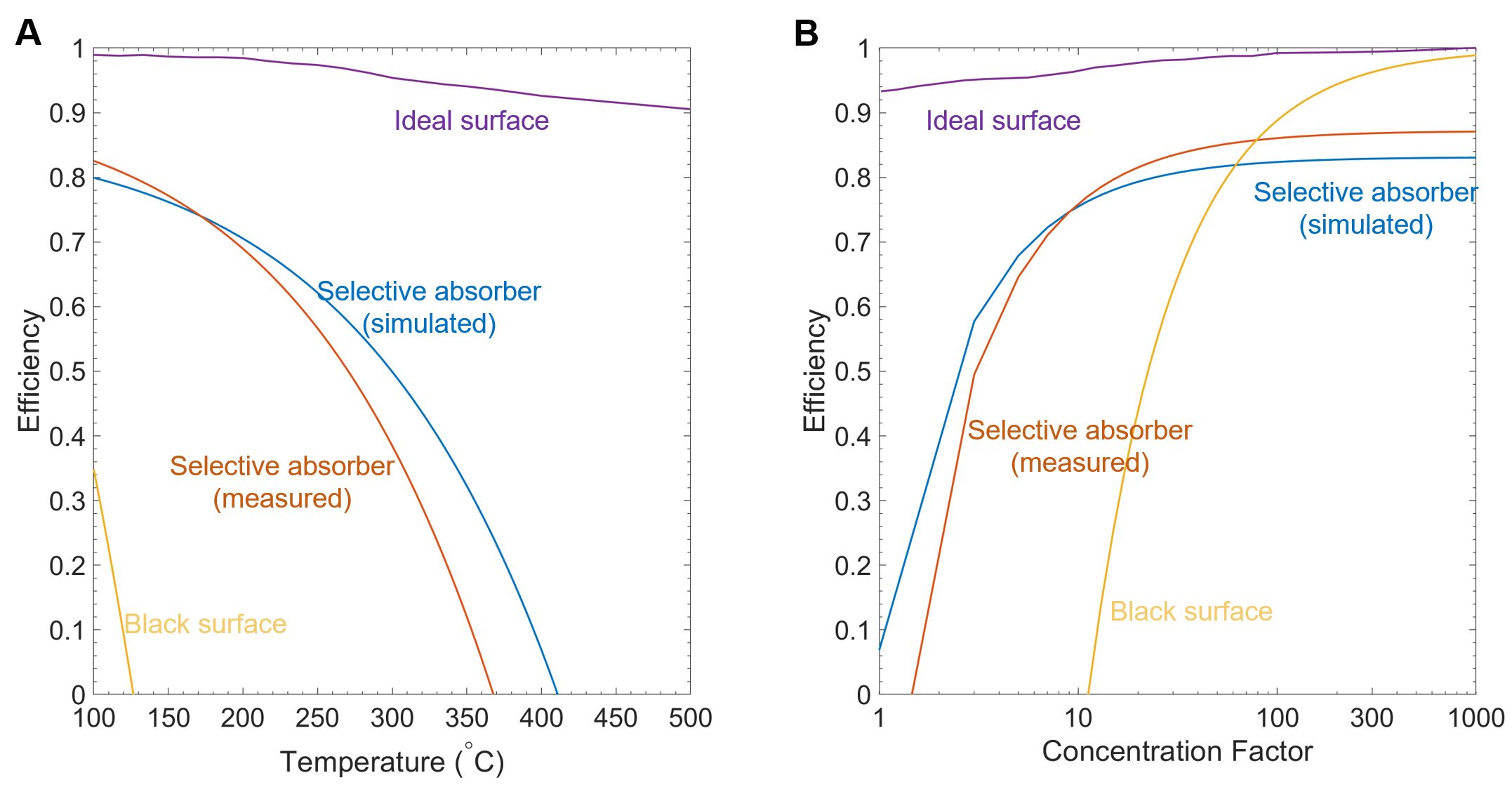}
\caption{ \label{fig:Fig_6} (\textbf{A}) The calculated solar to heat energy conversion efficiency of an ideal selective absorber, the multilayer solar absorber with radiative properties of measured or simulated, and a black surface as a function of absorber operational temperature, $T_\text{abs}$, under unconcentrated solar light; (\textbf{B}) Solar to heat conversion efficiency for abovementioned four absorber surfaces as a function of concentration factors, $CF$, at an absorber operational temperature of $T_\text{abs}$ = 400 $^{\circ}C$.
}
\end{figure}
In order to quantitatively evaluate the thermal performance of the designed solar absorber, the solar to heat conversion efficiency is theoretically analyzed here according to Eq. \ref{eq:e1}. Since the designed structure of solar absorber is demonstrated to be angular-insensitive as well as thermally stable up to 600 $^{\circ}C$, we use the reflectivity obtained from the UV/VIS/NIR spectrophotometer and FTIR spectrometer at room temperature (25 $^{\circ}C$) for efficiency calculation, as shown in Fig. \ref{fig:Fig_5}\textbf{A}. The $\alpha_{\lambda,abs}^{\prime}$ and $\epsilon_{\lambda,abs}^{\prime}$ in Eqs. \ref{eq:e2} and \ref{eq:e3} are assumed to be independent of temperature as observed in Fig. \ref{fig:Fig_5}\textbf{B}. The spectral integration for $\alpha_{\lambda,abs}^{\prime}$ and $\epsilon_{\lambda,abs}^{\prime}$ is performed over wavelengths regime from 0.3 $\mu$m to 16 $\mu$m, which cover the 97\% of the solar radiation wavelength regime. Only 5\% of the blackbody thermal radiation fall outside this defined spectral region for a 400 $^{\circ}C$ blackbody. 

Figure \ref{fig:Fig_6}\textbf{A} shows the solar to heat conversion as a function of absorber temperature, $T_\text{abs}$ under unconcentrated solar radiation for an ideal absorber surface, the multilayer solar absorber with spectral reflectivity taken from measurements or simulation, and a black surface (with unity absorptance over the entire wavelength of interest). The cut-off wavelength of an ideal solar absorber differs and is optimized at different operational temperatures to maximize the energy conversion efficiency, which indicates an upper limit of efficiency. The reflectivity of a black surface is zero over the entire wavelength region and shows no spectral selectivity.

It has been shown that the conversion efficiency of the solar absorber is 82.5\% and 79.9\% at an absorber temperature $T_\text{abs}$ = 100 $^{\circ}C$ with the simulated or measured spectral reflectivity. The efficiency drops gradually to zero at the stagnation temperature of 410 $^{\circ}C$ and 367 $^{\circ}C$ for the solar absorber using simulated and measured optical properties, where absorbed solar energy equals to blackbody re-emission energy (i.e., no solar thermal energy is actually harvested). The energy curves of solar absorber with measured and simulated radiative properties meet at 171 $^{\circ}C$. The fabricated solar absorber has a higher efficiency than the designed one below 171 $^{\circ}C$ because the reflectivity of the fabricated absorber is lower from 0.3 $\mu$m to 0.5 $\mu$m, where nearly half of the solar radiation distributes than the simulation values, which is shown in Fig. \ref{fig:Fig_5}\textbf{A}. Above 171 $^{\circ}C$, the efficiency of as designed solar absorber exhibits more advantages of the conversion efficiency than the fabricated one. It has a higher stagnation temperature due to its higher reflectivity in the mid-infrared region for enhanced minimization of thermal re-emission. As a reference, the black surface converts 34.8\% solar energy to heat at $T_\text{abs}$, and its efficiency goes down to zero at 126 $^{\circ}C$ very quickly, which further demonstrates the significance of wavelength selectivity in maximizing solar to heat energy conversion efficiency. On the other hand, the efficiency of the multilayer absorber is 17\% and 20\% lower than the ideal surface at $T_\text{abs}$ = 100 $^{\circ}C$ with measured optical properties and simulated radiative spectrum data, respectively. It mainly results from the larger thermal emittance in the mid-infrared regime. Additionally, the reflectivity of the selective absorber within the solar radiation spectrum is higher than that of the ideal one. The difference between the designed solar absorber and the ideal surface becomes even larger with the increasing of the absorber operational temperature. It is because that the cut-off wavelength of the ideal temperature is optimized at each temperature according to Wien's displacement law, while the cut-off wavelength of selective absorber keeps unchanged at 1.7 $\mu$m. Simultaneously, the reflectivity spectrum of the ideal surface changes more sharply at the cut-off point than the multilayer solar absorber, as shown in Fig. \ref{fig:Fig_5}\textbf{A}. Therefore, the geometry parameters of the multilayer solar absorber need to be optimized to make the cut-off wavelength perfectly matched to the operational temperature. 

The solar to heat energy conversion efficiency also varies with the concentration factors at CSP systems. Figure \ref{fig:Fig_6}\textbf{B} plots the efficiency as a function of concentration factor, $CF$, from 1 to 1,000, at an absorber temperature, $T_\text{abs}$ = 400 $^{\circ}C$, for a medium-temperature solar thermal application. The cut-off wavelength of the ideal selective absorber is optimized according to different $CF$s, and the corresponding efficiency goes up from 93.8\% to near unity when the $CF$ reaches 1,000, which indicates an upper limit for the selective solar absorber performance. The energy efficiency of both multilayer solar absorber with measured or simulated optical properties and black surface keeps going up with an increase of $CF$. The efficiency of solar absorber with measured reflective spectrum is lower than with simulated radiative properties below 10 suns, it is because the simulated data has a higher emittance than the measured one, however, when the $CF$ gets larger, the solar energy input will play a main role in the efficiency calculations and also the absorber with measured optical has lower reflectivity with solar radiation spectrum. For the black surface, its energy conversion efficiency becomes greater than zero at around 12 suns and climbs up approaching the thermal performance of ideal selective absorber when the $CF$ = 1,000. The solar radiation heat flux is much larger than the 600 $^{\circ}C$ blackbody thermal radiation under 1,000 concentrated sunlight. From Fig. \ref{fig:Fig_6}\textbf{B}, it can be concluded that the selective solar absorber has the advantage over the black surface below around 100 suns, while the selective solar absorber will fail when the $CF$ becomes larger.

\subsection{Thermal performance investigations under different concentration factors in a one-day sunlight cycle}

In order to demonstrate the absorption capability of the multilayer solar absorber under direct solar radiation with various $CF$, the temperature variations are simulated by solving thermal balance equation as expressed by the following \cite{ono2018self,yang2018dual}:

\begin{equation}\label{eq4}
\begin{split}
Q_\text{total} (T_\text{abs},T_\text{amb}) = &Q_\text{sun}(T_\text{abs})+Q_\text{amb} (T_\text{amb})\\
&-Q_\text{re-emit}(T_\text{abs})
\end{split}
\end{equation}

It is supposed that the backside of the solar absorber is thermally insulated (i.e., no thermal load is connected to absorber), so the heat transfer is considered only between the solar absorber and air. Here, $Q_\text{sun}$ is the heating power of solar absorber from solar radiation, $Q_\text{amb}$ is incident thermal radiation from ambient, $Q_\text{re-emit}$ stands for the heat flux through thermal re-emission from the solar absorber surface, and $Q_\text{total}$ is the net heating power of the solar absorber.

Solar radiation absorbed by the absorber, \textit{Q$_\text{sun}$}, is given by $Q_\text{sun}(T_\text{abs})$:

\begin{equation}\label{eq5}
Q_\text{sun}(T_\text{abs}) = A\cdot{CF}\int_0^\infty {\rm d}\lambda I_{\mathrm{AM} 1.5} (\lambda) \alpha (\lambda, \theta_\text{sun}, T_\text{abs})
\end{equation}  

Here, $A$ is the area of the solar absorber. $\alpha (\lambda, \theta_\text{sun}, T_\text{abs})$ is the temperature, wavelength-dependent and angular sensitive absorptance of the solar absorber, however, as discussed above, the absorptance of the designed solar absorber is angular and temperature-independent below 600 $^{\circ}C$. Hence, it is rational to take the measured data at room temperature into calculations.

The absorbed power of incident thermal radiation from atmosphere $Q_\text{amb} (T_\text{amb})$ can be expressed as follows:

\begin{equation}\label{eq6}
\begin{aligned}
Q_\text{amb}(T_\text{amb}) = A\int_0^\infty &{\rm d}\lambda I_\mathrm{BB} (T_\text{amb},\lambda) \alpha (\lambda, \theta, \phi, T_\text{abs})\\
&\times \epsilon (\lambda, \theta, \phi)
\end{aligned}
\end{equation}
where $I_\mathrm{BB} (T_\text{amb}, \lambda)$ = $2hc^{5}{\lambda^{-5}}$ $\exp( hc/\lambda k_{B}T-1)^{-1}$ defines the spectral radiance of a blackbody at a certain temperature \textit{T}, where \textit{h} is the Planck's constant, $k_{B}$ is the Boltzmann constant, and $\lambda$ is the wavelength. $\alpha(\lambda, \theta, \phi, T_\text{abs})$ = $\frac{1}{\pi}$$\int_0^{2\pi}{\rm d}\phi \int_0^{\pi/2} \epsilon_{\lambda}\cos\theta \sin \theta {\rm d}\theta$ is the temperature-dependent absorptance of solar absorber \cite{zhang2007nano}. Here, we take it as temperature-independent after demonstrations of high-temperature stability test. The emissivity of the air, $\epsilon (\lambda, \theta, \phi)$, is given by 1-$t (\lambda, \theta, \phi)$. $t (\lambda, \theta, \phi)$ is the transmittance value of atmosphere obtained from MODTRAN 4 \cite{berk1999modtran4}.

The heat flux through thermal re-emission from the solar absorber surface is determined as follows: 

\begin{equation}\label{eq7}
Q_\text{re-emit}(T_\text{abs}) = A\int_0^\infty {\rm d}\lambda I_\mathrm{BB} (T_\text{abs},\lambda) \epsilon (\lambda, \theta, \phi, T_\text{abs})
\end{equation}
where, $I_\mathrm{BB} (T_\text{abs}, \lambda)$ is the thermal radiation of a blackbody at a certain temperature. $\epsilon (\lambda, \theta, \phi, T_\text{abs})$ = $\alpha (\lambda, \theta, \phi, T_\text{abs})$ is the emissivity of the solar absorber according to Kirchhoff's law of thermal radiation \cite{robitaille2009kirchhoff}. 

The time-dependent temperature variations of the absorber can be obtained by solving the following equation:

\begin{equation}\label{eq8}
C_\text{abs} \frac{dT}{dt} = Q_\text{total}(T_\text{abs}, T_\text{amb})
\end{equation}
Since the multilayer structure of absorber is only 350 nm thick, it is reasonable to neglect its thermal resistance. Therefore, the heat capacitance of the absorber, $C_\text{abs}$, considers here equal to the heat capacitance of 300 $\mu$m silicon wafer on top of which the multilayer absorbers are fabricated.

\begin{figure}[ht]
\centering
\includegraphics[width=0.9\textwidth]{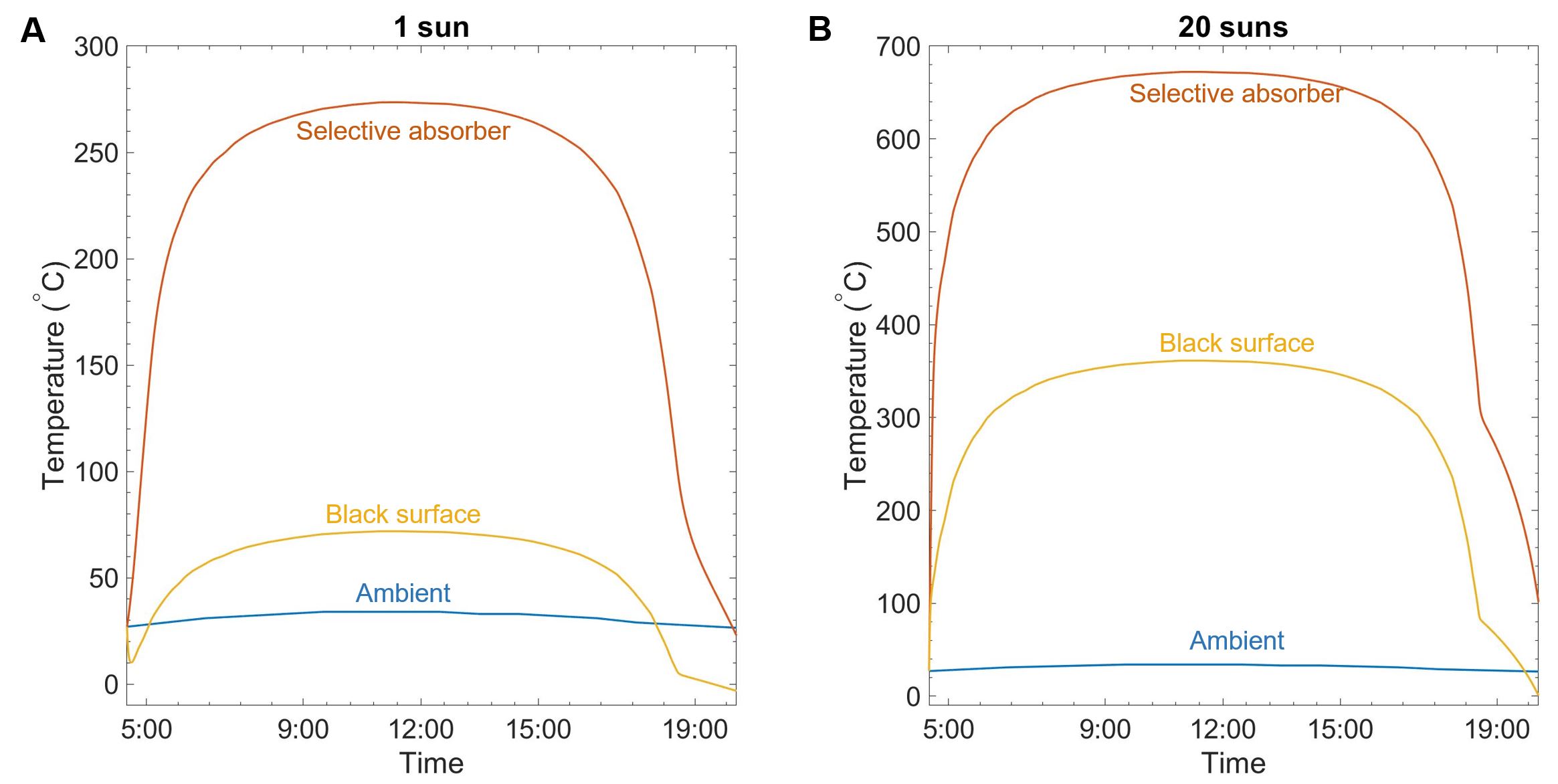}
\caption{ \label{fig:Fig_7} (\textbf{A}) and (\textbf{B}) Thermal performance of the selective absorber (orange curve) and the black surface (yellow curve) over a one-day sunlight cycle from sunrise (5:00 a.m.) to one hour after sunset (8:00 p.m.) at varying ambient temperature (blue curve) two different concentration factors (1 sun and 20 suns).
}
\end{figure}

The transient temperature fluctuations of the solar absorber under different concentration factors are simulated by solving Eq. \ref{eq8}, which is integrated over time to obtain the temperature evolution of solar absorber, as shown in Fig. \ref{fig:Fig_7}. For each simulation, the initial temperature of the solar absorber is assumed to be the same as the ambient temperature.

Figure \ref{fig:Fig_7} shows the transient temperature variations of the solar absorber under 1 sun (i.e. no optical concentration) and 20 suns, respectively, for fabricated selective absorber and black surface over a one-day sunlight cycle on July 10, 2018, in Boston, Massachusetts \cite{Weatherboston}, which is a typical summer climate. Using the ambient temperature \cite{Weatherboston} and the solar illumination data \cite{solarangleboston} of July 10, 2018, as the input data of Eq. \ref{eq4}, the temperature variations of the fabricated selective absorber and black surface are simulated from sunrise (5:00 a.m.) to one hour after sunset (8:00 p.m.). It can be observed that the highest temperature of the selective solar absorber is 273 $^{\circ}C$ and 672 $^{\circ}C$ for 1 sun and 20 suns, respectively, while the highest temperature of the black surface is 72 $^{\circ}C$ and 361 $^{\circ}C$ for 1 sun and 20 suns, respectively. It indicates a difference of their highest temperature between the selective and black solar absorber is 201 $^{\circ}C$ and 311 $^{\circ}C$ under 1 sun and 20 suns, respectively. It is obvious that the thermal performance of the selective solar absorber is overwhelmingly better than the black surface at any time under sunshine, and it reveals the significance of selective solar absorber in the CSP systems. It is worth to mention that, the temperature of the black surface starts to drop below the ambient temperature at about half-hour after the sunrise (5:30 a.m.) or half-hour before sunset (6:30 p.m.) and even drop below 0 $^{\circ}C$ at one hour after sunset (8:00 p.m.). It can be attributed to the radiative cooling properties of the black surface, considering that the black surface has unity emittance over the atmospheric transparent window (from 7.9 $\mu$m to 13 $\mu$m) and it has been discussed here \cite{yang2018dual,ono2018self}.

\begin{table}[!t]
\small
  \caption{\ Deposition methods and parameters of different layers of the designed absorber}
  \label{Tab3}
  \begin{tabular*}{\textwidth}{@{\extracolsep{\fill}}llllll}
    \hline
    Material & Deposition  & Layer  & Deposition  & Argon base  & Sputtering \\
             &  method    &  thickness & rate   &   pressure &  power \\
     &  &   (nm) &   ($\overset{\circ}A/s$) &  (10$^{-3}$ Torr) & (W) \\
    \hline
    SiO$_2$ coating & RF sputtering & 90 & 0.98 & 3.8 & 150\\
    Al$_2$O$_3$ top layer & RF sputtering & 70 & 1.09 & 3.8 & 15\\
    W top layer & DC sputtering & 10 & 0.72 & 0.8 & 150\\
    Al$_2$O$_3$ top layer & RF sputtering & 100 & 1.09 & 3.8 & 15\\
    W substrate & DC sputtering & 80 & 0.72 & 0.8 & 150\\
    \hline
  \end{tabular*}
\end{table}

\section{Methods}
\subsection{Sample fabrications and SEM topography characterizations}
The selective solar absorber samples are deposited on a 2'' silicon wafer using a home-built high vacuum magnetron sputtering system, details about this machine are described in this publication \cite{tian2019near}. The base pressure before the deposition is 3.4 $\times$ 10$^{-7}$ Torr. The fabricated parameters including deposition rate, base pressure, and sputtering power are specified in Table \ref{Tab3}. The SiO$_2$ subscript layer is grown from the silicon target using reactive radio frequency (RF) sputtering with argon and oxygen gas supplied during deposition. The total thickness of the fabricated absorber is 350 nm, which is shown in Fig. \ref{fig:Fig_3}\textbf{B}. The inset of Fig. \ref{fig:Fig_3}\textbf{B} shows a photo of the multilayer absorber sample is placed in a 2-inch single wafer carrier case. The black surface elucidates its high absorptance in the visible spectral region. The 80 nm W substrate layer is thick enough to block all the incident wavelengths of interest from 0.3 $\mu$m to 15 $\mu$m, so it can be treated as optically opaque. The cross-section topography of the absorber sample is characterized by SIGMA VP Field Emission-Scanning Electron Microscope (FE-SEM).

\subsection{Optical and radiative properties measurements}
Reflectance measurements in the ultraviolet, visible, and near-infrared regions are performed on an Agilent Cary 5000 spectrometer with a 150 mm PTFE based integrating sphere. Reflectance measurements are taken with a wavelength scan step of 1 nm at a normal incident angle and normalized to a labsphere spectralon reflectance standard. The near/mid-infrared reflectance measurements are completed on Jasco 6600 FTIR spectrometer together with a Pike 3 inches golden integrating sphere. The angle of the incident beam from the FTIR spectrometer is fixed at 12 $^{\circ}$. Spectra are taken at a scan rate of 64 with a wavelength resolution of 0.4 cm$^{-1}$. Details about FTIR spectrum measurements can be referred to Tian et al.'s recent article \cite{tian2019near}.

\subsection{High-temperature insensitivity test}
High-temperature stability tests are done in a tube oven at an argon protective atmosphere with an alumina tube of 5 cm diameter and 80 cm length. The tube is connected to a rotary vane vacuum pump, a vacuum gauge, and an argon tank at one end. Samples are placed in an alumina crucible boat (100 mm $\times$ 30 mm $\times$ 20 mm) and positioned in the center of the tube. First, the rotary van vacuum pump will vacuum down to 1.5 $\times$ 10$^{-2}$ Torr, then open the argon tank valve with a partial pressure of 70 psi and introduce argon into the alumina tube for 30 seconds. These processes are repeated three times for each test. The temperature controller is set to be 200 $^{\circ}C$, 400 $^{\circ}C$, 600 $^{\circ}C$, 800 $^{\circ}C$, and 1,000 $^{\circ}C$ for a 6-hour cycle. The reflectance measurements are carried out after each thermal treatment.

\section*{Funding}
National Science Foundation (NSF) (CBET-1941743); National Aeronautics and Space Administration (NNX15AK52A).\\
% \\
% OSA participates in \href{https://www.crossref.org/fundingdata/}{Crossref's Funding Data}, a service that provides a standard way to report funding sources for published scholarly research. To ensure consistency, please enter any funding agencies and contract numbers from the Funding section in Prism during submission or revisions. If exact wording for a funder is required, this may be added to the Acknowledgment section, even if it duplicates information in the funding information. 

\section*{Acknowledgments}
We thank Lijuan Qian and Dr. Gang Xiao for the help of the fabrication of selective solar absorbers and valuable discussion of experimental characterizations.

\section*{Disclosures}
The authors declare no conflict of interest.

\bibliography{Yanpei_Zheng}

\end{document}